# Bound states of the Dirac equation for the *PT*-symmetric generalized Hulthén potential by the Nikiforov-Uvarov method

**Harun Egrifes** [1,a] and **Ramazan Sever** [2,b]

[1] Department of Physics, Faculty of Science, Ege University, 35100 Izmir, Turkey

[2] Department of Physics, Faculty of Arts and Sciences, Middle East Technical University, 06531 Ankara, Turkey

## Abstract

The one-dimensional Dirac equation is solved for the *PT*-symmetric generalized Hulthén potential. The Nikiforov-Uvarov method which is based on solving the second-order linear differential equations by reduction to a generalized equation of hypergeometric type is used to obtain exact energy eigenvalues and corresponding eigenfunctions.

[a] E-mail : egrifes@sci.ege.edu.tr

[b] E-mail : sever@metu.edu.tr

## 1. Introduction

In the last few years there has been considerable work on non-Hermitian Hamiltonians. Among this kind of Hamiltonians, much attention has focused on the investigation of properties of so-called *PT*-symmetric Hamiltonians. Following the early studies of Bender and his co-workers [1], the *PT*-symmetry formulation has been successfully utilized by many authors [2-10]. The *PT*-symmetric but non-Hermitian Hamiltonians have real spectra whether the Hamiltonians are Hermitian or not. Non-Hermitian Hamiltonians with real or complex spectra have also been analyzed by using different methods [3-6,10-12]. Non-Hermitian but *PT*-symmetric models have applications in different fields, such as optics [13], nuclear physics [14], condensed matter [15], quantum field theory [16] and population biology [17].

The aim of the present Letter is to further pursue the development of *PT*-symmetry and to solve the one-dimensional Dirac equation for some complex systems. In view of the *PT*-symmetric formulation, we will apply the Nikiforov-Uvarov (NU) method [18] to solve the s-wave Dirac equation. We have presented exact bound states for a family of exponential-type potentials, i.e., generalized Hulthén potential which can be reduced to the standard Hulthén potential, Woods-Saxon potential and exponential-type screened Coulomb potential. This family of potentials have been applied with success to a number of different fields of physical systems. Using the quantization of the boundary condition of the states at the origin, Znojil [19] studied another form of the generalized Hulthén potential in non-relativistic and relativistic region. Dominguez-Adame [20] and Chetouani *et al* [21] also studied relativistic bound states of the standard Hulthén potential. On the other hand, Rao and Kagali [22] investigated relativistic bound states of the exponential-type screened Coulomb potential by means of the one-dimensional Klein-Gordon equation. However, it is well known that for the exponential-type screened Coulomb potential there is no explicit form of the energy expression of bound states for Schrödinger [23], Klein-Gordon [20] and also Dirac [24] equations.

In a recent work [25], we have presented the bound-state solutions of one-dimensional Klein-Gordon equation for *PT*-symmetric potentials with complexified generalized Hulthén potential. In this study we will be dealing with bound-state solutions of the one-dimensional Dirac equation for real and complex forms of generalized Hulthén potential. The organization of the present Letter is as follows. After a brief introductory discussion of the NU-method in Section 2, we obtain the bound-state energy eigenvalues for real and complex cases of generalized Hulthén potential and corresponding eigenfunctions in Section 3. In Section 4, we have presented the NU-method for exact bound states of exponential-type screened Coulomb potential [24]. As pointed out by Dutt *et al* [26], the screened Coulomb potential can very well be represented by an effective Hulthén potential [27]. Finally, conclusions and remarkable facts are discussed in the last section.

## 2. The Nikiforov-Uvarov method

Various special functions appear in the solution of many problems of relativistic and non-relativistic quantum mechanics. The differential equations whose solutions are the special functions can be solved by using the NU-method. This method is developed for constructing solutions of the general second-order linear differential equation which are special orthogonal polynomials [18]. It is well known that any given one-dimensional or radial Schrödinger equation can be written as a second-order linear differential equation. Therefore, to apply the NU-method one writes it in the generalized hypergeometric differential equation form

$$\phi''(z) + \frac{\tilde{\tau}(z)}{\sigma(z)}\phi'(z) + \frac{\tilde{\sigma}(z)}{\sigma^2(z)}\phi(z) = 0 \tag{1}$$

where $\sigma(z)$ and $\tilde{\sigma}(z)$ are polynomials, at most of second degree, and $\tilde{\tau}(z)$ is a polynomial, at most of first degree.

Using the transformation

$$\phi(z) = \varphi(z)\, y(z) \tag{2}$$

together with the equation determining the eigenvalues

$$\lambda = \lambda_n = -n\tau'(z) - \frac{n(n-1)}{2}\sigma''(z), \qquad (n = 0,1,2,...) \tag{3}$$

Eq.(1) can also be reduced to the following differential equation

$$\sigma(z)y''(z) + \tau(z)y'(z) + \lambda\, y(z) = 0. \tag{4}$$

This is a differential equation of hypergeometric type, whose polynomial solutions are given by Rodrigues relation [18]

$$y_n(z) = \frac{B_n}{\rho(z)} \frac{d^n}{dz^n}\left[\sigma^n(z)\rho(z)\right] \tag{5}$$

where

$$\frac{\varphi'(z)}{\varphi(z)} = \frac{\pi(z)}{\sigma(z)} \tag{6}$$

$$\tau(z) = \tilde{\tau}(z) + 2\pi(z) \tag{7}$$

$$\pi(z) = \frac{\sigma'(z) - \tilde{\tau}(z)}{2} \pm \sqrt{\left(\frac{\sigma'(z) - \tilde{\tau}(z)}{2}\right)^2 - \tilde{\sigma}(z) + k\sigma(z)} \tag{8}$$

$$k = \lambda - \pi'(z). \tag{9}$$

Since $\pi(z)$ has to be a polynomial of degree at most one, in Eq.(8) the expression under the square root must be the square of a polynomial of first degree. This is possible only if its discriminant is zero. Hence we obtain an equation, in the quadratic form, for $k$. After determining $k$, we have $\pi(z)$ from Eq.(8), and then $\varphi(z)$, $\tau(z)$ and $\lambda$, respectively.

### 3. Exact bound-state solutions of the generalized Hulthén potential

It is well known that the exact solutions of the Dirac equation play an important role in relativistic quantum mechanics. Thus, considerable efforts have been spent in recent years towards obtaining the exact solutions of other types of relativistic wave equations for certain potentials of physical interest [28-34]. Having given the brief review of the NU-method above, let us now consider the one-dimensional time-independent Dirac equation with any given interaction potential $V(x)$ in the vector coupling scheme ( in units where $\hbar = c = 1$ ) [24]

$$\left\{ i \frac{d}{dx} \begin{pmatrix} 0 & -1 \\ 1 & 0 \end{pmatrix} + [E - V(x)] \begin{pmatrix} 0 & 1 \\ 1 & 0 \end{pmatrix} - m \begin{pmatrix} 1 & 0 \\ 0 & 1 \end{pmatrix} \right\} \psi(x) = 0. \tag{10}$$

The spinor wave function $\psi(x)$ has two components. We denote the upper (lower) component by $\phi$ ($\theta$). We will be dealing with bound-state solutions, i.e., the wave function vanishes at infinity.

We choose the one-dimensional vector potential

$$V_q(x) = -V_0 \frac{e^{-\alpha x}}{1 - q e^{-\alpha x}} \tag{11}$$

which is called generalized Hulthén potential [25]. The deformation parameter $q$ determines the shape of the potential. It is worth mentioning here that, for some specific $q$ values this potential transforms to the well-known types : such as for $q = 0$ to the exponential potential, for $q = 1$ to the standard Hulthén potential and for $q = -1$ to the Woods-Saxon potential. When $\alpha \to 0$, the potential is close to the origin

$$V_q(x) \approx \frac{V_0}{q-1} + \frac{V_0}{(q-1)^2} \alpha x, \tag{12}$$

and behaves like a linear potential with a constant shift, $\frac{V_0}{q-1}$, where $\alpha$ denotes the range parameter and $V_0$ denotes the coupling constant.

For the potential function given by Eq.(11), Eq.(10) decompose into :

$$\left( i\frac{d}{dx} + E + V_0 \frac{e^{-\alpha x}}{1 - q e^{-\alpha x}} \right) \phi_q(x) = m \theta_q(x) \qquad (13)$$

$$\left( -i\frac{d}{dx} + E + V_0 \frac{e^{-\alpha x}}{1 - q e^{-\alpha x}} \right) \theta_q(x) = m \phi_q(x). \qquad (14)$$

The coupled differential equations allow bound-state solutions requiring the vanishing of the wave function at infinity. In terms of $\phi_q(x)$ and $\theta_q(x)$ the spinor wave function is normalized, so that $\phi_q(x)$ and $\theta_q(x)$ are square-integrable functions.

It is well known the fact that the above set of coupled equations can be reduced to a second-order differential equation. Substituting Eq.(13) into Eq.(14), one obtains the following Schrodinger-like second-order differential equation for the upper component

$$\phi_q''(x) + \left[ \tilde{E} + V_1 \frac{e^{-2\alpha x}}{(1 - q e^{-\alpha x})^2} + V_2 \frac{e^{-\alpha x}}{1 - q e^{-\alpha x}} \right] \phi_q(x) = 0 \qquad (15)$$

where

$$\tilde{E} = E^2 - m^2, \quad V_1 = V_0^2 + i q\alpha V_0, \quad V_2 = i\alpha V_0 + 2EV_0. \qquad (16)$$

By defining a new variable $s = e^{-\alpha x}$, this equation is reduced to the generalized equation of hypergeometric type which is given by Eq.(1)

$$\phi_q''(s) + \frac{1-qs}{s-qs^2}\phi_q'(s) + \frac{1}{(s-qs^2)^2}\left[ \left(\gamma_q - q\beta - q^2\varepsilon^2\right)s^2 + \left(\beta + 2q\varepsilon^2\right)s - \varepsilon^2 \right]\phi_q(s) = 0 \quad (17)$$

for which

$$\tilde{\tau}_q(s) = 1 - qs, \quad \sigma_q(s) = s - qs^2, \quad \tilde{\sigma}_q(s) = \left(\gamma_q - q\beta - q^2\varepsilon^2\right)s^2 + \left(\beta + 2q\varepsilon^2\right)s - \varepsilon^2,$$

$$\gamma_q = i\frac{qV_0}{\alpha} + \frac{V_0^2}{\alpha^2}, \quad \beta = i\frac{V_0}{\alpha} + \frac{2V_0 E}{\alpha^2}, \quad \varepsilon^2 = -\frac{1}{\alpha^2}\left(E^2 - m^2\right), \qquad (18)$$

with real $\varepsilon^2 > 0$ ($E^2 < m^2$) for bound states. Substituting $\sigma_q(s), \tilde{\tau}_q(s)$ and $\tilde{\sigma}_q(s)$ into Eq.(8), one finds

$$\pi_q(s) = -\frac{qs}{2} \pm \frac{1}{2}\sqrt{\left[q^2 - 4\left(\gamma_q - q\beta - q^2\varepsilon^2\right) - 4qk\right]s^2 + 4\left[k - \left(\beta + 2q\varepsilon^2\right)\right]s + 4\varepsilon^2}. \quad (19)$$

The constant parameter $k$ can be determined from the condition that the expression under the square root has a double zero, i.e., $k = \beta \pm \sqrt{q^2 - 4\gamma_q}\,\varepsilon$. We then obtain the following possible forms of $\pi_q(s)$:

$$\pi_q(s) = -\frac{qs}{2} \pm \begin{cases} \frac{1}{2}\left[(v - 2q\varepsilon)s + 2\varepsilon\right] & \text{for} \quad k = \beta + v\,\varepsilon \\[2mm] \frac{1}{2}\left[(v + 2q\varepsilon)s - 2\varepsilon\right] & \text{for} \quad k = \beta - v\,\varepsilon \end{cases} \qquad (20)$$

where $v = \sqrt{q^2 - 4\gamma_q} = q - i\frac{2V_0}{\alpha}$. The polynomial $\pi_q(s)$ is chosen such that the function $\tau_q(s)$ given by Eq.(7) will have a negative derivative [18]. This condition is satisfied by

$$\tau_q(s) = 1 + 2\varepsilon - (2q + v + 2q\varepsilon)s \qquad (21)$$

which corresponds to

$$\pi_q(s) = \varepsilon - \frac{1}{2}\left[q + (v + 2q\varepsilon)\right]s . \qquad (22)$$

Then, we have another constant, $\lambda = k + \pi'_q(s)$, written as

$$\lambda = \beta - \frac{1}{2}(v + q) - (v + q)\,\varepsilon . \qquad (23)$$

Thus, substituting $\lambda, \tau'_q(s)$ and $\sigma''_q(s)$ into Eq.(3), the exact energy eigenvalues of the generalized Hulthén potential are determined as

$$E_{nq} = \frac{V_0}{2q} \pm i\kappa_n(q,\alpha,V_0)\sqrt{\frac{1}{4q^2} - \frac{m^2}{V_0^2 + \kappa_n^2(q,\alpha,V_0)}} \qquad (24)$$

where $\kappa_n(q,\alpha,V_0) = q\alpha(n+1) - iV_0$.

To find the function $y(s)$, which is the polynomial solution of hypergeometric-type equation, we multiply Eq.(4) by an appropriate function $\rho(s)$ so that it can be written in self-adjoint form [18]

$$(\sigma\,\rho\,y')' + \lambda\,\rho\,y = 0 . \qquad (25)$$

Here $\rho(s)$ satisfies the differential equation $(\sigma\,\rho)' = \tau\,\rho$ which yields

$$\rho_q(s) = s^{2\varepsilon}(1 - qs)^{v/q} . \qquad (26)$$

We thus obtain the eigenfunctions of hypergeometric-type equation from the Rodrigues relation given by Eq.(5) in the following form:

$$y_{nq}(s) = B_{nq}\,s^{-2\varepsilon}(1 - qs)^{-v/q}\frac{d^n}{ds^n}\left[s^{n+2\varepsilon}(1 - qs)^{n+(v/q)}\right]. \qquad (27)$$

The functions $y_{nq}(s)$ are, up to a numerical factor, the Jacobi polynomials $P_n^{(2\varepsilon, \nu/q)}(z)$ with $z = 1 - 2qs$. By substituting $\pi_q(s)$ and $\sigma_q(s)$ in Eq.(6), one can find the other factor of the upper component giving

$$\varphi_q(s) = s^\varepsilon (1-qs)^{(\nu+q)/2q}. \tag{28}$$

As stated in Eq.(2), the wave function is constructed as a combination of two independent parts. In this case the upper spinor component can be determined as

$$\phi_{nq}(s) = \varphi_q(s)\, y_{nq}(s)$$
$$= C_{nq}\, s^\varepsilon (1-qs)^{(\nu+q)/2q}\, P_n^{(2\varepsilon, \nu/q)}(1-2qs) \tag{29}$$

with $s = e^{-\alpha x}$. Notice that $\phi_{nq}(x)$ decreases exponentially as $x \to \infty$, being square-integrable and thus representing a truly bound-state.

By using the differential and recursion properties of the Jacobi polynomials [35], the lower spinor component can also be derived from Eq.(13) as

$$m\theta_{nq}(s) = C_{nq}\, s^\varepsilon (1-qs)^{(\nu+q)/2q} \left\{ \left[ E + i\alpha\left(n + \varepsilon + \frac{\nu}{q} + 1\right)\right] + \left(V_0 + i\alpha\frac{\nu+q}{2}\right)\frac{s}{1-qs}\right\} P_n^{(2\varepsilon, \nu/q)}(1-2qs) -$$
$$- C_{nq}\, i\alpha\left(n + 2\varepsilon + \frac{\nu}{q} + 1\right) s^\varepsilon (1-qs)^{(\nu+q)/2q}\, P_n^{\left(2\varepsilon, \frac{\nu}{q}+1\right)}(1-2qs) \tag{30}$$

An inspection of the energy expression given by Eq.(24) shows that, for any given $\alpha$, the spectrum consist of complex eigenvalues depending on $q$. Now let us discuss the conditions for having real energy eigenvalues. As we will see the role played by the range parameter $\alpha$ is very crucial in this regard. To this end we consider the complexified forms of the generalized Hulthén potential.

### 3.1 Non-Hermitian PT-symmetric generalized Hulthén potential

A Hamiltonian is said to be *PT*-symmetric when $[H, PT] = 0$, where $P$ is the parity operator, i.e., $P: V(x) \to V(-x)$, and $T$ is the complex conjugation operator, i.e., $T: i \to -i$, namely, *PT*-symmetry condition for a given potential $V(x)$ reads

$$[V(-x)]^* = V(x). \tag{31}$$

Now let us consider the case, namely, at least one of the potential parameters is complex. If $\alpha$ is a pure imaginary parameter, i.e. $\alpha \to i\alpha$, such potentials are written as a complex function

$$V_q(x) = \frac{V_0}{q^2 - 2q\cos(\alpha x) + 1}\left[q - \cos(\alpha x) + i\sin(\alpha x)\right], \tag{32}$$

which is *PT*-symmetric but non-Hermitian. It is worthwhile here to point out that, such as a complex periodic potential having *PT*-symmetry of the form $V(x) = i \sin^{2n+1}(x)$ $(n = 0,1,2,...)$ which exhibit real band spectra was discussed in detail by Bender *et al*. [36]. The complex potential (32) has real spectra given by

$$E_{nq} = \frac{V_0}{2q} \mp (q\alpha(n+1) - V_0) \sqrt{\frac{1}{4q^2} - \frac{m^2}{V_0^2 - (q\alpha(n+1) - V_0)^2}} \quad (33)$$

if and only if $4q^2 m^2 \leq V_0^2 - (q\alpha(n+1) - V_0)^2$. Moreover, this restriction which gives the critical coupling value, namely, $\frac{q\alpha}{2}(n+1) + \frac{2qm^2}{\alpha(n+1)} \leq V_0$, leads to the result

$$\frac{V_0 - q\alpha - \sqrt{V_0^2 - 4q^2 m^2}}{q\alpha} \leq n \leq \frac{V_0 - q\alpha + \sqrt{V_0^2 - 4q^2 m^2}}{q\alpha}, \quad (34)$$

i.e., there exist finite number of eigenvalues.

The corresponding wave functions are

$$\phi_{nq}(s) = C_{nq} \, s^{i\varepsilon} (1-qs)^{(a+q)/2q} \, P_n^{(2i\varepsilon, a/q)}(1-2qs), \quad (35)$$

$$m\theta_{nq}(s) = C_{nq} \, s^{i\varepsilon} (1-qs)^{(a+q)/2q} \left\{ \left[ E - \alpha\left(n + i\varepsilon + \frac{a}{q} + 1\right) \right] + \left(V_0 - \alpha\frac{a+q}{2}\right) \frac{s}{1-qs} \right\} P_n^{(2i\varepsilon, a/q)}(1-2qs) +$$

$$+ C_{nq} \, \alpha\left(n + 2i\varepsilon + \frac{a}{q} + 1\right) s^{i\varepsilon} (1-qs)^{(a+q)/2q} \, P_n^{\left(2i\varepsilon, \frac{a}{q}+1\right)}(1-2qs), \quad (36)$$

where $a = q - 2\frac{V_0}{\alpha}$ and $s = e^{-i\alpha x}$.

Figs.1.(a) and (b) shows the variation of the ground-state level (i.e. $n = 0$) as a function of the coupling constant $V_0$ for different positive and negative shape parameters, and $1/\alpha = \lambda_c$, where $\lambda_c = \hbar/mc = 1/m$, denotes the Compton wavelength of the Dirac particle. As it can be seen from Fig.1.(a), while *PT*-symmetric non-Hermitian generalized Hulthén potential generates real and positive bound states for positive shape parameters, it generates real and negative bound states for the same value of $\alpha$ when $q < 0$ ( Fig.1.(b) ). Obviously, for any given $V_0$, as seen from Figs.2.(a) and 2.(b), all possible eigenstates have positive (negative) eigenenergies if the parameter $q$ is positive (negative). It is almost notable that there are some crossing points of the relativistic energy eigenvalues for some $\alpha$ values. One can see from Eq. (34) that why there is a limitation on the number of bound states of *PT*-symmetric generalized Hulthén potential. With the arbitrary choice of the parameters adopted

in ploting Fig.2.(a), it can be seen that the range parameter $\alpha$ varies as $m \leq \alpha \leq 4m$ for $n = 0$, $0.5 m \leq \alpha \leq 2 m$ for $n = 1$, $0.33 m \leq \alpha \leq 1.33 m$ for $n = 2$ etc.

### 3.2 Pseudo-Hermiticity and PT-symmetry

It is interesting to note that when all three parameters $V_0$, $q$ and $\alpha$ are imaginary at the same time, i.e., $V_0 \to i V_0$, $q \to i q$ and $\alpha \to i\alpha$, the potential transforms to the form

$$V_q(x) = \frac{V_0}{q^2 - 2q \sin(\alpha x) + 1}[q - \sin(\alpha x) - i\cos(\alpha x)]. \tag{37}$$

This form of the potential has a $\pi/2$ phase difference with respect to the potential given by Eq.(32). Recently, Mostafazadeh [8] has shown that the potentials of this form are $P$-pseudo-Hermitian and claimed that the $\eta$-pseudo-Hermiticity, $\eta H \eta^{-1} = H^+$, is the necessary condition for having real spectrum, where $\eta$ is referred to as a Hermitian linear automorphism. For a non-Hermitian Hamiltonian $H$ to have a complete set of biorthonormal eigenvectors, the necessary and sufficient condition is to possess an invertible linear operator $O$ such that $H$ is $\eta$-pseudo-Hermitian, where $\eta = O O^+$. Ahmed [37] has suggested an explicit form for a Hermitian linear automorphism, $\eta = e^{-\theta p}$, $p = -i d/dx$, which affects an imaginary shift of the coordinate : $\eta x \eta^{-1} = x + i\theta$.

If we replace $x$ by $\left(\frac{\pi}{2\alpha} - x\right)$, we have $\sin(\alpha x) \to \cos(\alpha x)$ and also $\cos(\alpha x) \to \sin(\alpha x)$. Thus we obtain $P V_q(x) P^{-1} = V_q^*(x)$ for the complex potential given by Eq.(37) where the Hermitian, linear and invertible operator $\eta$ is the the parity operator $P$, which acts on the position operator as $P x P^{-1} = \frac{\pi}{2\alpha} - x$ [38]. Hence, the complex version of the generalized Hulthén potential possesses $P$-pseudo-Hermiticity. Under joint action of spatial reflection $\left(P : x \to \frac{\pi}{2\alpha} - x\right)$ and time reversal $(T : i \to -i)$, we obtain $PT V_q(x) (PT)^{-1} = V_q(x)$ for the complex version of the potential function given by Eq.(11). Therefore, we can say that the complex potential given by Eq.(37) also holds $PT$-symmetry. It has exact real spectra

$$E_{nq} = \frac{V_0}{2q} \mp (V_0 - q\alpha(n+1))\sqrt{\frac{1}{4q^2} - \frac{m^2}{V_0^2 - (V_0 - q\alpha(n+1))^2}} \tag{38}$$

if and only if $4q^2m^2 \leq V_0^2 - (V_0 - q\alpha(n+1))^2$. Now referring back to Eqs.(29) and (30), the corresponding wave functions $\psi_{nq}(s)$ are identified in the form

$$\phi_{nq}(s) = C_{nq}\, s^{i\varepsilon}\, (1-iqs)^{(a+q)/2q}\, P_n^{(2i\varepsilon,\, a/q)}(1-2iqs), \tag{39}$$

$$m\theta_{nq}(s) = C_{nq}\, s^{i\varepsilon}\, (1-iqs)^{(a+q)/2q} \left\{ \left[ E - \alpha\left(n + i\varepsilon + \frac{a}{q} + 1\right)\right] + i\left(V_0 - \alpha\frac{a+q}{2}\right)\frac{s}{1-iqs} \right\} P_n^{(2i\varepsilon,\, a/q)}(1-2iqs) +$$

$$+ C_{nq}\, \alpha\left(n + 2i\varepsilon + \frac{a}{q} + 1\right) s^{i\varepsilon}\, (1-iqs)^{(a+q)/2q}\, P_n^{\left(2i\varepsilon,\, \frac{a}{q}+1\right)}(1-2iqs), \tag{40}$$

where $s = e^{-i\alpha x}$.

### 4. The solution of the generalized Hulthén potential for $q = 0$

In the previous section we have obtained the bound-state solutions of the generalized Hulthén potential with $q \neq 0$ and presented the explicit form of the eigenvalues and the spinor wave functions. Now we turn our attention to the $q = 0$ case. In this case, the potential given by Eq.(11) can be regarded as a screened one-dimensional Coulomb potential [24]. Note that, for $q = 0$, there is no explicit form of the energy expression of bound states for Schrödinger [23], Klein-Gordon [22] and also Dirac equations [24]. Hence we have to resolve it, which is similar to the screened Coulomb potential [24].

For $q = 0$ the generalized equation of hypergeometric type which is given by Eq.(17) becomes

$$\phi''(s) + \frac{1}{s}\phi'(s) + \frac{1}{s^2}\left[\frac{V_0^2}{\alpha^2}s^2 + \beta s - \varepsilon^2\right]\phi(s) = 0 \tag{41}$$

and the corresponding $\pi(s)$ is determined as

$$\pi(s) = \pm \begin{cases} i\delta s + \varepsilon & \text{for } k = \beta + 2i\delta\varepsilon \\ i\delta s - \varepsilon & \text{for } k = \beta - 2i\delta\varepsilon \end{cases} \tag{42}$$

where $\delta = \frac{V_0}{\alpha}$. Following a procedure similar to the previous case, when $\pi(s) = -i\delta s + \varepsilon$ is chosen for $k = \beta - 2i\delta\varepsilon$,

$$\tau(s) = 1 + 2\varepsilon - 2i\delta s,\; \lambda = \beta - i\delta - 2i\delta\varepsilon,\; \varphi(s) = s^\varepsilon\, e^{-i\delta s} \tag{43}$$

could be obtained. Substituting $\sigma(s)$ and $\tau(s)$, together with $\lambda$, into Eq.(4) yields

$$s\, y''(s) + [(1+2\varepsilon) - 2i\delta s]\, y'(s) - (i\delta + 2i\delta\varepsilon - \beta)\, y(s) = 0. \tag{44}$$

This equation can also be reduced to the standard Whittaker differential equation [39]. Thus, the solutions vanishing at infinity can be written in terms of confluent hypergeometric function as follows :

$$y(s) = {}_1F_1\left(\frac{1}{2}+\varepsilon+i\frac{\beta}{2\delta};1+2\varepsilon;2i\delta s\right). \tag{45}$$

So, the acceptable solution for the upper component is found to be

$$\phi(s) = \varphi(s)y(s) = A\, s^{\varepsilon} e^{-i\delta s}\, {}_1F_1\left(\frac{1}{2}+\varepsilon+i\frac{\beta}{2\delta};1+2\varepsilon;2i\delta s\right) \tag{46}$$

A being a normalization constant. Eq.(13) gives the lower spinor component in terms of the upper as

$$m\theta(s) = A\,(-i\alpha\varepsilon + E)\, s^{\varepsilon} e^{-i\delta s}\, {}_1F_1\left(\varepsilon+i\frac{E}{\alpha};2\varepsilon+1;2i\delta s\right)+$$

$$+ A\frac{2V_0}{2\varepsilon+1}\left(\varepsilon+i\frac{E}{\alpha}\right)s^{\varepsilon+1} e^{-i\delta s}\, {}_1F_1\left(\varepsilon+i\frac{E}{\alpha}+1;2\varepsilon+2;2i\delta s\right). \tag{47}$$

The corresponding bound-state eigenvalue spectra can be obtained only by using numerical methods [24,40].

### 5. Conclusions

We have seen that the s-wave Dirac equation for generalized vector Hulthén potential can be solved exactly. The relativistic bound-state energy spectrum and the corresponding wave functions have been obtained by the NU-method. Some interesting results including complex *PT*-symmetric and pseudo-Hermitian versions of the generalized Hulthén potential have also been discussed. We have already mentioned that we have found some simple relations among the potential parameters for bound states. We show that it is possible to obtain relativistic bound states of complex quantum mechanical formulation.

## References

[1] C.M. Bender, S. Boettcher, Phys.Rev.Lett. 80 (1998) 5243.

   C.M. Bender, S. Boettcher, P.N. Meisenger, J.Math.Phys. 40 (1999) 2201.

[2] C.M. Bender, G.V. Dunne, J.Math.Phys. 40 (1999) 4616.

   F.M. Fernandez, R. Guardiola, J. Ros, M. Znojil, J.Phys. A : Math. Gen. 31 (1998) 10105.

   G.A. Mezincescu, J.Phys. A : Math. Gen. 33 (2000) 4911.

   E. Delabaere, D.T. Trinh, J.Phys. A : Math. Gen. 33 (2000) 8771.

   M. Znojil, M. Tater, J.Phys. A : Math. Gen. 34 (2001) 1793.

   C.M. Bender, G.V. Dunne, P.N. Meisenger, M. Simsek, Phys.Lett. A 281 (2001) 311.

   Z. Ahmed, Phys.Lett. A 282 (2001) 343.

   Z. Ahmed, Phys.Lett. A 284 (2001) 231.

   P. Dorey, C. Dunning, R. Tateo, J.Phys. A : Math. Gen. 34 (2001) 5679.

   B. Bagchi, C. Quesne, Mod.Phys.Lett. A 16 (2001) 2449.

   K.C. Shin, J.Math.Phys. 42 (2001) 2513.

   K.C. Shin, Commun.Math.Phys. 229 (2002) 543.

   C.K. Mondal, K. Maji, S.P. Bhattacharyya, Phys.Lett. A 291 (2001) 203.

   G.S. Japaridze, J.Phys. A : Math. Gen. 35 (2002) 1709.

   C.S. Jia, P.Y. Lin, L.T. Sun, Phys.Lett. A 298 (2002) 78.

   C.S. Jia, S.C. Li, Y.Li, L.T. Sun, Phys.Lett. A 300 (2002) 115.

   C.S. Jia, L.Z. Yi, Y. Sun, J.Y. Liu, L.T. Sun, Mod.Phys.Lett. A 18 (2003) 1247.

   C.S. Jia, Y. Li, Y. Sun, J.Y. Liu, L.T. Sun, Phys.Lett. A 311 (2003) 115.

[3] F. Cannata, G. Junker, J. Trost, Phys.Lett. A 246 (1998) 219.

[4] A. Khare, B.P. Mandal, Phys.Lett. A 272 (2000) 53.

[5] B.Bagchi, C. Quesne, Phys.Lett. A 273 (2000) 285.

[6] Z. Ahmed, Phys.Lett. A 282 (2001) 343.

[7] L. Solombrino, J.Math.Phys. 43 (2002) 5439.

[8] A. Mostafazadeh, J.Math.Phys. 43 (2002) 205.

   A. Mostafazadeh, J.Math.Phys. 43 (2002) 2814.

   A. Mostafazadeh, J.Math.Phys. 43 (2002) 3944.

[9] M.Znojil, Phys.Lett. A 264 (1999) 108.

[10] B. Bagchi, C. Quesne, Phys.Lett. A 300 (2002) 18.

[11] C.M. Bender, M. Berry, P.N. Meisenger, V.M. Savage, M. Simsek, J.Phys. A : Math. Gen. 34 (2001) L31.

[12] C.M. Bender, E.J Weniger, J.Math.Phys. 42 (2001) 2167.

C.M. Bender, S. Boettcher, H.F. Jones, P.N. Meisenger, M. Simsek, Phys.Lett. A 291 (2001) 197.

[13] S.K. Moayedi, A. Rostami, Eur.Phys.J. B 36 (2003) 359.

[14] D. Baye, G. Levai, J.M. Sparenberg, Nucl.Phys. A 599 (1996) 435.

R.N. Deb, A. Khare, B.D. Roy, Phys.Lett. A 307 (2003) 215.

[15] N. Hatano, D.R. Nelson, Phys.Rev.Lett. 77 (1996) 570.

N. Hatano, D.R. Nelson, Phys.Rev.B 56 (1997) 8651.

[16] C.M. Bender, K.A. Milton, V.M. Savage, Phys.Rev. D 62 (2000) 085001.

C. Bernard, V.M. Savage, Phys.Rev. D 64 (2001) 085010.

[17] D.R.Nelson, N.M. Shnerb, Phys.Rev.E 58 (1998) 1383.

[18] A.F. Nikiforov, V.B. Uvarov, Special Functions of Mathematical Physics, Birkhauser, Basel, 1988.

[19] M. Znojil, J.Phys. A : Math. Gen. 14 (1981) 383.

[20] F. Dominguez-Adame, Phys.Lett. A 136 (1989) 175.

[21] L. Chetouani, L Guechi, A. Lecheheb, T.F. Hammann, A. Messouber, Physica A 234 (1996) 529.

[22] N.A. Rao, B.A. Kagali, Phys.Lett. A 296 (2002) 192.

[23] S. Flügge, Practical Quantum Mechanics, Springer, Berlin, 1974.

[24] F. Dominguez-Adame, A. Rodriguez, Phys.Lett. A 198 (1995) 275.

V.M. Villalba and W.Greiner, Phys.Rev.A 67 (2003) 052707.

[25] M. Simsek, H. Egrifes, J.Phys. A : Math. Gen. 37 (2004) 4379.

[26] R. Dutt, K. Chowdhury, Y.P Varshni, J.Phys. A : Math. Gen. 18 (1985) 1379.

[27] R.L. Greene, C. Aldrich, Phys.Rev. A 14 (1976) 2363.

[28] Y.Nogami, F.M. Toyama, Phys.Rev. A 57 (1998) 93.

F.M. Toyama, Y. Nogami, Phys.Rev. A 59 (1999) 1056.

[29] N.A. Rao, B.A. Kagali, V. Sivramkrishna, Int.J.Mod.Phys. A 17 (2002) 4739.

B.A. Kagali, N.A. Rao, V. Sivramkrishna, Mod.Phys.Lett. A 17 (2002) 2049.

[30] S.H. Dong, Z.Q. Ma, Phys.Lett. A 312 (2003) 78.

[31] O. Mustafa, J.Phys. A : Math. Gen. 36 (2003) 5067.

[32] A.D. Alhaidari, Phys.Rev. A 65 (2002) 042109.

A.D. Alhaidari, J.Phys. A : Math. Gen. 37 (2004) 5805.

A.D. Alhaidari, Phys.Lett. A 322 (2004) 72.

A.D. Alhaidari, Phys.Lett. A 326 (2004) 58.

[33] G. Chen, Phys.Lett. A 328 (2004) 116.

   G. Chen, Z.D. Chen, Z.M. Lou, Phys.Lett. A 331 (2004) 374.

[34] Y.F. Diao, L.Z. Yi, C.S. Jia, Phys.Lett. A 332 (2004) 157.

   L.Z. Yi, Y.F. Diao, J.Y. Liu, C.S. Jia, Phys.Lett. A 333 (2004) 212.

[35] W. Magnus, F. Oberhettinger and R.P. Soni, Formulas and Theorems for the Special Functions of Mathematical Physics, 3rd. ed., Springer, Berlin, 1966.

[36] C.M. Bender, G.V. Dunne, P.N. Meisinger, Phys.Lett.A 252 (1999) 272.

[37] Z. Ahmed, Phys.Lett. A 290 ( 2001) 19.

[38] C.S. Jia, P.Y. Lin, L.T. Sun, Phys.Lett. A 298 ( 2002) 78.

   C.S. Jia, Y. Sun, Y.Li, Phys.Lett. A 305 (2002) 231.

[39] M. Abromowitz and I. Stegun, Handbook of Mathematical Functions with Formulas, Graphs and Mathematical Tables, Dover, New York, 1964.

[40] W. Greiner, Relativistic Quantum Mechanics, Springer-Verlag, Berlin, 1990.

Figure 1.(a)

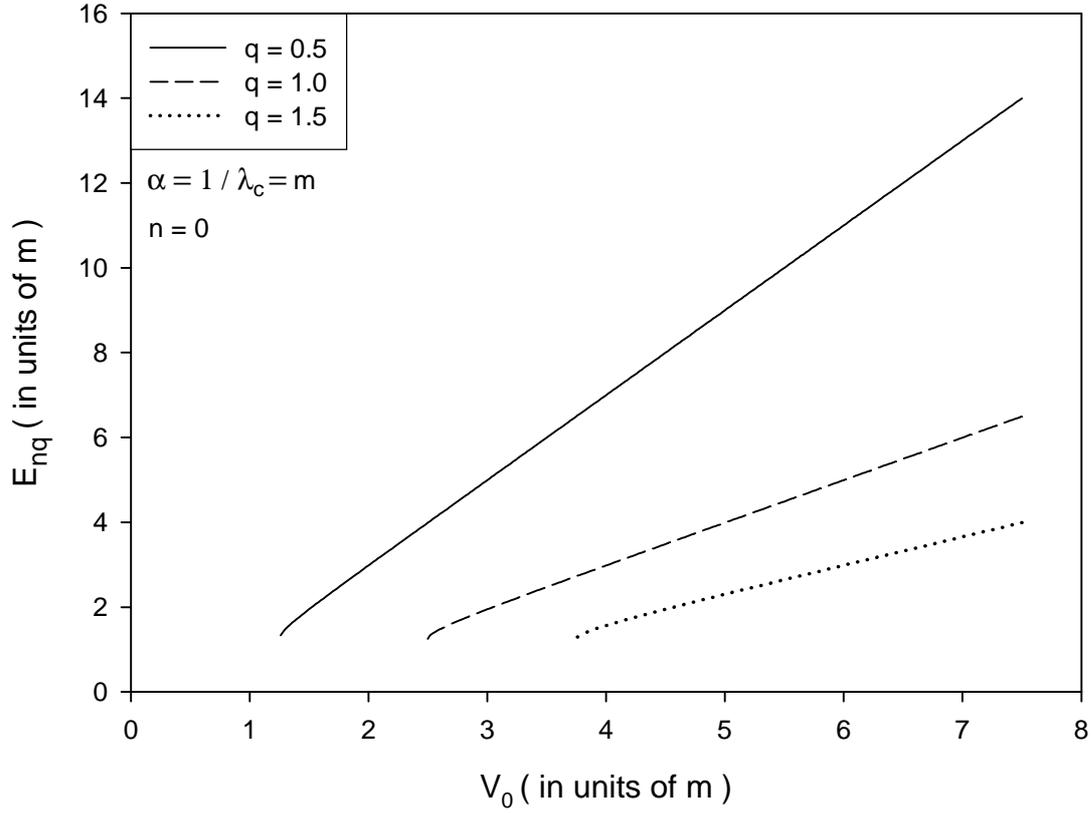

**Fig.1.(a) :** The variation of the ground-state energy of a Dirac particle, which is moving in *PT*-symmetric potential given by Eq.(32), as a function of the coupling constant $V_0$ for three different positive shape parameters.

Figure 1.(b)

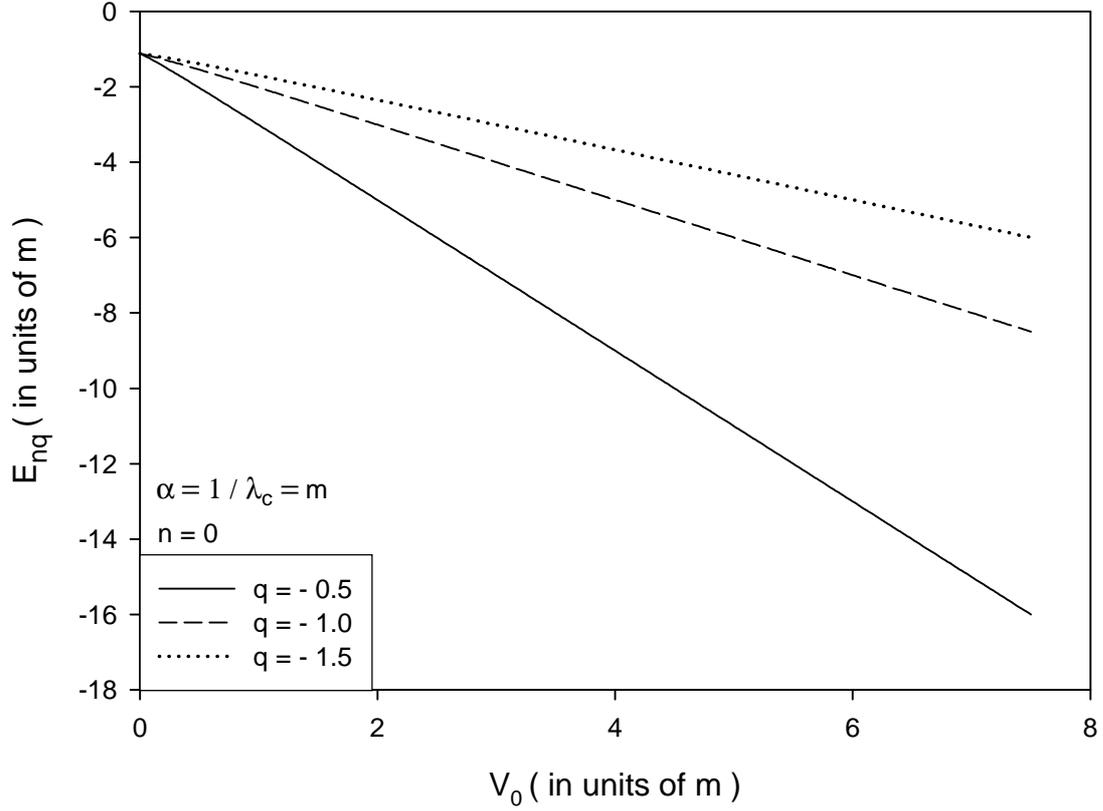

**Fig.1.(b) :** The variation of the ground-state energy of a Dirac particle, which is moving in *PT*-symmetric potential given by Eq.(32), as a function of the coupling constant $V_0$ for three different negative shape parameters.

Figure 2.(a)

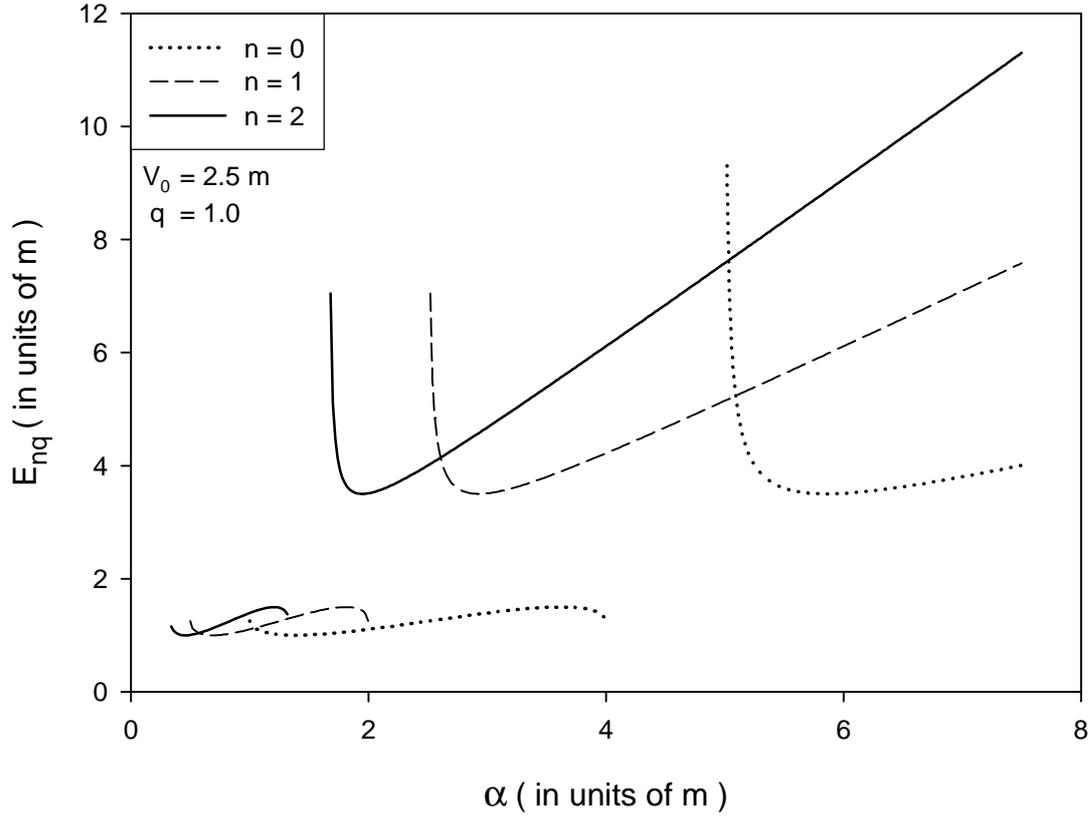

**Fig.2.(a) :** The variation of the energy eigenvalues of a Dirac particle, which is moving in *PT*-symmetric potential given by Eq.(32), as a function of the range parameter $\alpha$ for the positive shape parameter ($q = 1.0$). The curves are plotted for the first three values of the vibrational quantum number $n$.

Figure 2.(b)

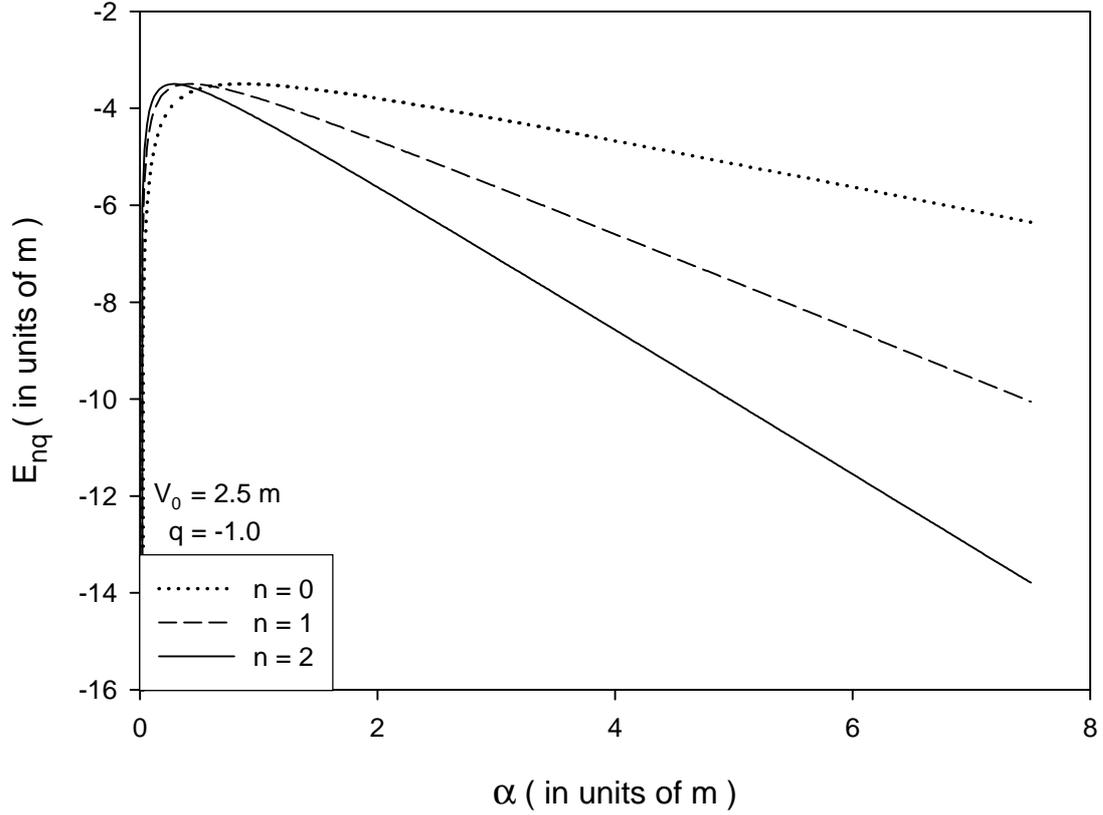

**Fig.2.(b) :** The variation of the energy eigenvalues of a Dirac particle, which is moving in *PT*-symmetric potential given by Eq.(32), as a function of the range parameter $\alpha$ for the negative shape parameter ($q = -1.0$). The curves are plotted for the first three values of the vibrational quantum number $n$.